\documentclass[aps,preprint,floats,epsf,epsfig,nofootinbib,letter]{revtex4-1}

\def\bea{\begin{eqnarray}}
\def\eea{\end{eqnarray}}

\def\ov{\overline}

\def\pr{{Phys. Rev.}~}

\def\lsim{ {\ \lower-1.2pt\vbox{\hbox{\rlap{$<$}\lower5pt\vbox{\hbox{$\sim$}
}}}\ } }
\def\gsim{ {\ \lower-1.2pt\vbox{\hbox{\rlap{$>$}\lower5pt\vbox{\hbox{$\sim$}
}}}\ } }

\begin{document}
\begin{flushright}
{\small
CYCU-HEP-10-21 }
\end{flushright}

\title{\Large \bf $1^{++}$ Nonet Singlet-Octet Mixing Angle, Strange Quark Mass, and Strange Quark Condensate}

\author{\vspace*{0.3cm} \sc Kwei-Chou Yang}\email{Email: kcyang@cycu.edu.tw}

\bigskip
\affiliation{\vspace*{0.3cm}  Department of Physics,
Chung Yuan Christian University, Chung-Li 320, Taiwan \vspace*{2.3cm} }


\begin{abstract}{\small
Two strategies are taken into account to determine the $f_1(1420)$-$f_1(1285)$ mixing angle $\theta$. (i) First, using the Gell-Mann-Okubo mass formula together with the $K_1(1270)$-$K_1(1400)$ mixing angle $\theta_{K_1}=(-34\pm 13)^\circ$ extracted from the data for ${\cal B}(B\to K_1(1270) \gamma), {\cal B}(B\to K_1(1400) \gamma), {\cal B}(\tau\to K_1(1270) \nu_\tau)$, and ${\cal B}(\tau\to K_1(1420) \nu_\tau)$, gave $\theta = (23^{+17}_{-23})^\circ$. (ii) Second, from  the study of the ratio for $f_1(1285) \to \phi\gamma$ and $f_1(1285) \to \rho^0\gamma$ branching fractions, we have a two-fold solution $\theta=(19.4^{+4.5}_{-4.6})^\circ$ or $(51.1^{+4.5}_{-4.6})^\circ$. Combining these two analyses, we thus obtain $\theta=(19.4^{+4.5}_{-4.6})^\circ$.
We further compute the strange quark mass and strange quark condensate from the analysis of the  $f_1(1420)$-$f_1(1285)$ mass difference QCD sum rule, where the operator-product-expansion series is up to dimension six and to ${\cal O}(\alpha_s^3, m_s^2 \alpha_s^2)$ accuracy. Using the average of the recent lattice results and the $\theta$ value that we have obtained as inputs, we get $\langle \bar{s} s \rangle/ \langle \bar{u} u\rangle =0.41 \pm 0.09$.}
\end{abstract}
\small
\maketitle

\section{Introduction}

The $f_1(1285)$ and $f_1(1420)$ mesons with quantum number $J^{PC}= 1^{++}$ are the members of the $1^3P_1$ states in the quark model language, and are mixtures of the pure octet $f_8$ and singlet $f_1$, where the mixing is characterized by the mixing angle $\theta$. The BaBar results for the upper bounds of $B^-\to f_1(1285)K^-,f_1(1420)K^-$ were available recently
\cite{Burke}. The relative ratio of these two modes is highly sensitive to $\theta$ \cite{Cheng:2007mx}.
On the other hand, in the two-body $B$ decay involving the $K$ meson in the final state, the amplitude receives large corrections from the chiral enhancement $a_6$ term which is inversely proportional to the strange-quark mass.
The quark mass term mixes left- and right-handed quarks in the QCD Lagrangian. The spontaneous breaking of chiral symmetry from $SU(3)_L \times SU(3)_R$ to $SU(3)_V$ is further broken by the quark masses $m_{u,d,s}$ when the baryon number is added to the three commuting conserved quantities $Q_u, Q_d$, and $Q_s$, respectively, the numbers of $q-\bar q$ quarks for $q=u,d$, and $s$. The nonzero quark condensate which signals dynamical symmetry breaking is the important parameter in QCD sum rules \cite{SVZ}, while the magnitude of the strange quark mass can result in the flavor symmetry breaking in the quark condensate. In an earlier study $\langle \bar{s} s\rangle /\langle \bar{u} u\rangle \sim 0.8 <1$ was usually taken. However, very recently the Jamin-Lange approach \cite{Jamin:2001fw} together with the lattice result for $f_{B_s}/f_B$ \cite{Aoki:2010yq} and also the Schwinger-Dyson equation approach \cite{Williams:2007ef} can give a central value larger than 1.

In this paper,  we shall embark on the study of the $f_1(1420)$ and $f_1(1285)$ mesons to determine the mixing angle $\theta$, strange quark mass, and strange quark condensate. In Sec. \ref{sec:theta}, we shall present detailed discussions on the determination of the mixing angle $\theta$. Substituting the $K_1(1270)$-$K_1(1400)$ mixing angle, which was extracted from the $B\to K_1 \gamma$ and $\tau\to K_1 \nu_\tau$ data, to the Gell-Mann-Okubo mass formula, we can derive the value of $\theta$. Alternatively, from  the analysis of the decay ratio for $f_1(1285) \to \phi\gamma$ and $f_1(1285) \to \rho^0\gamma$, we have a more accurate estimation for $\theta$. In Sec. \ref{sec:sr} we shall obtain the mass difference QCD sum rules for the $f_1(1420)$ and $f_1(1285)$ to determine the magnitude of the strange quark mass. From the sum rule analysis, we obtain the constraint ranges for $m_s$ and $\theta$ as well as for $\langle \bar{s} s\rangle$.
Many attempts have been made to compute $m_s$ using QCD sum rules and finite energy sum rules \cite{Dominguez:2007my,Chetyrkin:2005kn,Narison:2005ny,Kambor:2000dj,Pich:1999hc,Colangelo:1997uy,Jamin:1994vr}. The running strange quark mass in the $\overline{\rm MS}$ scheme at a scale of  $\mu\approx$ 2 GeV is $m_s=101^{+29}_{-21}$ MeV given in the particle data group (PDG) average \cite{PDG}. More precise lattice estimates have been recently obtained as $m_s(2\, {\rm GeV})=92.2(1.3)$ MeV in \cite{McNeile:2010ji}, $m_s(2\, {\rm GeV})=96.2(2.7)$ MeV in \cite{Aoki:2010dy}, and $m_s(2\, {\rm GeV})=95.1(1.1)(1.5)$ MeV in \cite{Durr:2010aw}. These lattice results agree with strange scalar/pseudoscalar sum rule results which are $m_s \simeq 95(15)$ MeV. In the present study, we study the $m_s$ from a new frame, the $f_1(1420)$-$f_1(1285)$ mass difference sum rule, which may result in larger uncertainties due to the input parameters. Nevertheless, it can be a crosscheck compared with the previous studies. Further using the very recent lattice result for $m_s(2~{\rm GeV})=93.6\pm 1.0$~MeV as the input, we obtain an estimate for the strange quark condensate.

\section{Singlet-octet mixing angle $\theta$ of the $1^{++}$ nonet}\label{sec:theta}

\subsection{Definition}\label{sec:def}
~~~ In the quark model, $a_1(1260)$, $f_1(1285)$, $f_1(1420)$, and $K_{1A}$ are classified in $1^{++}$ multiplets, which, in terms of spectroscopic notation $n^{2S+1}L_J$, are $1^3P_1$ $p$-wave mesons. Analogous to $\eta$ and $\eta^\prime$,
because of SU(3) breaking effects, $f_1(1285)$ and $f_1(1420)$ are
the mixing states of the pure octet $f_8$ and singlet $f_1$,
 \begin{eqnarray}
 |f_1(1285)\rangle = |f_1\rangle\cos\theta+|f_8\rangle\sin\theta,
 \qquad |f_1(1420)\rangle =
 -|f_1\rangle\sin\theta+|f_8\rangle\cos\theta \,.
 \end{eqnarray}
In the present paper, we adopt
\begin{eqnarray}
 f_1 &=& \frac{1}{\sqrt{3}}(\bar u u + \bar dd + \bar ss), \\
f_8 &=& \frac{1}{\sqrt{6}}(\bar uu + \bar dd -2 \bar ss),
\end{eqnarray}
where there is a relative sign difference between the $\bar{s}s$ contents of $f_1$ and $f_8$ in our
convention.  From the Gell-Mann-Okubo mass formula, the mixing angle $\theta$
satisfies
\begin{equation}
\cos^2\theta =
 \frac{3m_{f_1(1285)}^2 -\left(4m_{K_{1A}}^2-m_{a_1}^2\right)}
{3\left(m_{f_1(1285)}^2-m_{f_1(1420)}^2\right)}\,,\label{eq:GMOkubo}
\end{equation}
where
\begin{eqnarray}
 m_{K_{1A}}^2 &=& \langle K_{1A}| {\cal H}| K_{1A}\rangle
 =  m_{K_1(1400)}^2 \cos^2\theta_{K_1} + m_{K_1(1270)}^2 \sin^2\theta_{K_1} \,,
 \end{eqnarray}
with ${\cal H}$
being the Hamiltonian. Here $\theta_{K_1}$ is the $K_1(1400)$-$K_1(1270)$ mixing angle. The sign of the mixing angle $\theta$ can be determined from the mass relation \cite{PDG}
\begin{equation}
\tan\theta =\frac{4m_{K_{1A}}^2-m_{a_1}^2-3m_{f_1(1420)}^2}
{3 m_{18}^2} \,,\label{eq:GMOkubo2}
\end{equation}
where $m_{18}^2=\langle f_1|{\cal H} |f_8 \rangle \simeq (m_{a_1}^2-m_{K_{1A}}^2)2\sqrt{2}/3 <0$, we find $\theta>0$.
Because of the strange and nonstrange light quark mass differences, $K_{1A}$ is not the mass eigenstate and it can mix with $K_{1B}$, which is one of the members in the $1^1P_1$ multiplets.
From the convention in  \cite{Suzuki:1993yc} (see also discussions in \cite{Yang:2005gk,Yang:2007zt}), we write the two physical states $K_1(1270)$ and $K_1(1400)$ in the following relations:
 \begin{eqnarray}
 \label{eq:mixing}
 |K_1(1270)\rangle &=& |K_{1A}\rangle\sin\theta_K+ |K_{1B}\rangle\cos\theta_K, \nonumber \\
 |K_1(1400)\rangle &=& |K_{1A}\rangle\cos\theta_K - |K_{1B}\rangle\sin\theta_K.
 \end{eqnarray}
The mixing angle was found to be
$|\theta_{K_1}|\approx 33^\circ, 57^\circ$ in \cite{Suzuki:1993yc} and $\approx \pm37^\circ, \pm58^\circ$ in \cite{ChengDAP}. A similar range $35^\circ\lsim |\theta_{K_1}| \lsim 55^\circ$ was obtained in \cite{Goldman}. The sign ambiguity for $\theta_{K_1}$ is due to the fact that one can add arbitrary phases to $|\bar K_{1A}\rangle$ and $|\bar K_{1B}\rangle$. This sign ambiguity can be removed by fixing the signs of decay constants $f_{K_{1A}}$ and $f_{K_{1B}}^\perp$, which are defined by
  \begin{eqnarray}\label{eq:k1a}
 \langle 0 |\bar \psi\gamma_\mu \gamma_5 s |\bar K_{1A}(P,\lambda)\rangle
 &=& -i \, f_{K_{1A}}\, m_{K_{1A}}\,\epsilon_\mu^{(\lambda)},
 \end{eqnarray}
\begin{eqnarray}\label{eq:k1b}
 \langle 0 |\bar \psi\sigma_{\mu\nu}s |\bar K_{1B}(P,\lambda)\rangle
 &=& i f_{K_{1B}}^\perp
 \,\epsilon_{\mu\nu\alpha\beta} \epsilon_{(\lambda)}^\alpha
 P^\beta,
 \end{eqnarray}
where $\epsilon^{0123}=-1$ and $\psi \equiv u$ {\rm or} $d$. Following the convention in
 \cite{Yang:2007zt}, we adopt $f_{K_{1A}}>0$, $f_{K_{1B}}^\perp>0$, so that
$\theta_{K_1}$ should be negative to
account for the observable ${\cal B}(B\to K_1(1270) \gamma) \gg {\cal B}(B\to
K_1 (1400) \gamma)$ \cite{ChengKgamma,Hatanaka:2008xj}.
Furthermore, from the data of $\tau\to K_1(1270)\nu_\tau$ and
$K_1(1400)\nu_\tau$ decays together with the sum rule results for the $K_{1A}$ and $K_{1B}$ decay constants, the mixing angle $\theta_{K_1}=(-34\pm 13)^\circ$ was obtained in \cite{Hatanaka:2008xj}. Substituting this value into (\ref{eq:GMOkubo}), we then obtain $\theta^{\rm quad}=(23^{+17}_{-23})^\circ$ \cite{YangFF}, i.e., $\theta^{\rm quad}=0^\circ - 40^\circ$ \footnote{ Replacing the meson mass squared $m^2$ by $m$ throughout  (\ref{eq:GMOkubo}), we obtain $\theta^{\rm lin}= (23^{+17}_{-23})^\circ$. The difference is negligible. Our result can be compared with that using $\theta_{K_1}=-57^\circ$ into (\ref{eq:GMOkubo}), one has  $\theta^{\rm quad}=52^\circ$.}.

\subsection{The determination of $\theta$}

Experimentally, since $K^*\ov K$ and $K\ov K\pi$ are the dominant modes of $f_1(1420)$, whereas $f_0(1285)$ decays mainly to the $4\pi$ states, this suggests that the quark content is primarily $s\bar s$ for $f_1(1420)$ and $n\bar n=(u\bar{u} + d\bar{d})/\sqrt{2}$ for $f_1(1285)$. Therefore, the mixing relations can be rewritten to exhibit the $n\bar n$ and $s\bar s$ components which decouple for the ideal mixing angle $\theta_i=\tan^{-1}(1/\sqrt{2})\simeq 35.3^\circ$. Let $\bar{\alpha} =\theta_i -\theta$, we rewrite these two states in the flavor basis \footnote{In PDG \cite{PDG}, the mixing angle is defined as $\alpha =\theta -\theta_i +\pi/2$. Comparing it with our definition, we have $\alpha=\pi/2 -\bar\alpha$.},
\begin{eqnarray}
f_1(1285) &=& \frac{1}{\sqrt{2}} (\bar{u}u+\bar{d}d) \cos\bar\alpha + \bar{s}s\, \sin\bar\alpha \,, \nonumber\\
f_1(1420) &=& \frac{1}{\sqrt{2}} (\bar{u}u+\bar{d}d) \sin\bar\alpha - \bar{s}s\, \cos\bar\alpha \,.
\end{eqnarray}

Since the $f_1(1285)$ can decay into $\phi\gamma$, we know that  $f_1(1285)$ has the $s \bar{s}$ content and  $\theta$ deviates from its ideal mixing value. To have a more precise estimate for $\theta$,  we study the ratio of $f_1(1285) \to \phi\gamma$ and $f_1(1285) \to \rho^0\gamma$  branching fractions. Because the electromagnetic (EM) interaction Lagrangian is given by
\begin{eqnarray}
{\cal L}_I &=& -A_{\rm EM}^\mu (e_u \bar u\gamma_\mu u + e_d \bar d\gamma_\mu d + e_s \bar s\gamma_\mu s )  \nonumber\\
           &=& -A_{\rm EM}^\mu \left( (e_u +e_d)\frac{\bar u\gamma_\mu u +  \bar d\gamma_\mu d}{2}
            + (e_u -e_d)\frac{\bar u\gamma_\mu u - \bar d\gamma_\mu d}{2}
            + e_s \bar s\gamma_\mu s \right),
\end{eqnarray}
with $e_u=2/3 e, e_d =-1/3 e$, and $e_s=-1/3 e$ being the electric charges of  $u,d$, and $s$ quarks, respectively,
we obtain
\begin{eqnarray}
\frac{{\cal B}(f_1(1285) \to \phi\gamma)}{{\cal B}(f_1(1285) \to \rho^0\gamma)}
&=&
\left(\frac{\langle\phi|e_s\bar{s}\gamma_\mu s|f_1(1285)\rangle}
{\langle\rho|(e_u-e_d)(\bar{u}\gamma_\mu u -\bar{d}\gamma_\mu d)/2|f_1(1285)\rangle}\right)^2
\underbrace{\left(\frac{m_{f_1}^2-m_\phi^2}{m_{f_1}^2 - m_\rho^2}\right)^3}_{\rm phase\ factor}
\nonumber\\
&=& \underbrace{\left(\frac{-e/3}{2e/3+e/3}\right)^2}_{\rm EM\ factor}
 \left(\frac{\langle\phi|\bar{s}\gamma_\mu s|f_1(1285)\rangle}
 {\langle\rho|(\bar{u}\gamma_\mu u -\bar{d}\gamma_\mu d)/2|f_1(1285)\rangle}\right)^2
\underbrace{\left(\frac{m_{f_1}^2-m_\phi^2}{m_{f_1}^2 - m_\rho^2}\right)^3}_{\rm phase\ factor}
\nonumber\\
&\approx&  \frac{4}{9} \left(\frac{m_\phi f_\phi}{m_\rho f_\rho}\right)^2
\tan^2\bar\alpha  \left(\frac{m_{f_1}^2-m_\phi^2}{m_{f_1}^2 - m_\rho^2}\right)^3 \,,
\end{eqnarray}
where $f_1\equiv f_1(1285)$, and $f_\phi$ and $f_\rho$ are the decay constants of $\phi$ and $\rho$, respectively. Here we have taken the single-pole approximation \footnote{The following approximation was used in \cite{Close:1997nm}:
\begin{eqnarray}
\frac{\langle\phi|\bar{s}\gamma_\mu s|f_1(1285)\rangle}
 {\langle\rho|(\bar{u}\gamma_\mu u -\bar{d}\gamma_\mu d)/2|f_1(1285)\rangle}
 \approx  2\tan\bar\alpha \,. \nonumber
\end{eqnarray}
}:
\begin{eqnarray}
\frac{\langle\phi|\bar{s}\gamma_\mu s|f_1(1285)\rangle}
 {\langle\rho|(\bar{u}\gamma_\mu u -\bar{d}\gamma_\mu d)/2|f_1(1285)\rangle}
&\approx& \frac{m_\phi f_\phi g_{f_1\phi\phi}}
 {m_\rho f_\rho g_{f_1\rho\rho}/\sqrt{2}} \frac{\sin\bar\alpha}{\cos\bar\alpha /\sqrt{2}} \nonumber\\
 & \approx& \frac{m_\phi f_\phi}{m_\rho f_\rho}\times 2\tan\bar\alpha \,.
\end{eqnarray}
Using $f_\rho =209\pm 1$ MeV, $f_\phi=221\pm 3$ MeV \cite{Beneke:2003zv}, and the current data ${\cal B}(f_1(1285) \to \phi\gamma)=(7.4\pm2.6)\times 10^{-4}$ and ${\cal B}(f_1(1285) \to \rho^0 \gamma) =(5.5\pm1.3)\%$ \cite{PDG} as inputs, we obtain $\bar\alpha =\pm(15.8^{+4.5}_{-4.6})^\circ$, i.e., two fold solution $\theta=(19.4^{+4.5}_{-4.6})^\circ$ or $(51.1^{+4.5}_{-4.6})^\circ$. Combining with the analysis $\theta=(0 \sim 40)^\circ$ given in Sec. \ref{sec:def}, we thus find that $\theta=(19.4^{+4.5}_{-4.6})^\circ$ is much preferred and can explain experimental observables well.

\section{Mass of the strange quark}\label{sec:sr}
~~~We proceed to evaluate the strange quark mass from the mass difference sum rules of the $f_1(1285)$ and $f_1(1420)$ mesons. We consider the following two-point correlation functions,
\begin{eqnarray}\label{eq:green-axial}
\Pi_{\mu\nu} (q^2) &=& i\int d^4x e^{iqx} \langle 0|{\rm T} (j_\mu(x)
j_\nu^{\dag} (0))|0\rangle =-\Pi_1(q^2) g_{\mu \nu} +\Pi_2(q^2)
q_\mu q_\nu \,,
 \\
\Pi^\prime_{\mu\nu} (q^2) &=& i\int d^4x e^{iqx} \langle 0|{\rm T} (j^\prime_\mu(x)
j_\nu^{\prime\dag} (0))|0\rangle =-\Pi^\prime_1(q^2) g_{\mu \nu} +\Pi^\prime_2(q^2)
q_\mu q_\nu \,.
\end{eqnarray}
The interpolating currents satisfying the relations:
\begin{eqnarray}
    \langle 0 |j^{(\prime)}_\mu(0) |f_1^{(\prime)}(P,\lambda)\rangle
 =-if_{f_1^{(\prime)}} m_{f_1^{(\prime)}} \epsilon_\mu^{(\lambda)},
  \label{eq:axial-decayconstant}
\end{eqnarray}
are
\begin{eqnarray}
j_\mu &=& \cos\theta j^{(1)}_\mu + \sin\theta j^{(8)}_\mu \,,
 \\
j^\prime_\mu &=& -\sin\theta j^{(1)}_\mu + \cos\theta j^{(8)}_\mu \,,
\end{eqnarray}
where
\begin{eqnarray}
j^{(1)}_\mu &=& \frac{1}{\sqrt{3}}
(\bar{u}\gamma_\mu\gamma_5 u + \bar{d}\gamma_\mu\gamma_5 d +\bar{s}\gamma_\mu\gamma_5 s) \,,
 \\
j^{(8)}_\mu &=& \frac{1}{\sqrt{6}}
(\bar{u}\gamma_\mu\gamma_5 u + \bar{d}\gamma_\mu\gamma_5 d - 2\bar{s}\gamma_\mu\gamma_5 s) \,,
\end{eqnarray}
and we have used the short-hand notations for $f_1 \equiv f_1(1285)$ and $f_1^\prime \equiv f_1(1420)$.
In the massless quark limit, we have $\Pi_1=q^2\Pi_2$ and $\Pi^\prime_1=q^2\Pi^\prime_2$ if one neglects the axial-vector anomaly\footnote{Considering the anomaly, the singlet axial-vector current is satisfied with
$$\partial^\mu j^{(1)}_\mu=\frac{1}{\sqrt{3}}(m_u \bar uu+ m_d\bar dd+ m_s\bar ss)+\frac{3\alpha_s}{4\pi}G\tilde{G}$$.}. Here we focus on $\Pi_1^{(\prime)}$ since it
receives contributions only from axial-vector ($^3P_1$) mesons,
whereas $\Pi_2^{(\prime)}$ contains effects from pseudoscalar mesons.
The lowest-lying $f_1^{(\prime)}$ meson contribution can be approximated via the dispersion
relation as
\begin{eqnarray}
\frac{m_{f_1^{(\prime)}}^2 f_{f_1^{(\prime)}}^2}{m_{f_1^{(\prime)}}^2-q^2}
 = \frac{1}{\pi}\int^{s_0^{f^{(\prime)}}}_0 ds
\frac{{\rm Im} \Pi_1^{(\prime){\rm OPE}}(s)}{s-q^2} \,, \label{eq:dispersion}
\end{eqnarray}
where $\Pi_1^{(\prime){\rm OPE}}$ is the QCD operator-product-expansion (OPE)
result of $\Pi_1^{(\prime)}$ at the quark-gluon level \cite{Yang:2007zt}, and $s_0^{f_1^{(\prime)}}$ is the threshold of the higher resonant states. Note that the subtraction terms on the right-hand side of (\ref{eq:dispersion}), which are polynomials in $q^2$, are neglected since they have no contributions after performing the Borel transformation.
The four-quark condensates are expressed as
\begin{eqnarray}\label{eq:factorization}
\langle 0|\bar q \Gamma_i \lambda^a q \bar q \Gamma_i \lambda^a
q|0\rangle =- a_2\frac{1}{16N_c^2}{\rm Tr}(\Gamma_i\Gamma_i) {\rm
Tr}(\lambda^a \lambda^a) \langle \bar qq\rangle^2 \,,
\end{eqnarray}
where $a_2=1$ corresponds to the vacuum saturation approximation. In the present work, we have $\Gamma=\gamma_\mu$ and $\gamma_\mu\gamma_5$, for which we allow the variation $a_2 = -2.9\sim 3.1$  \cite{Narison:2005ny,Ackerstaff:1998yj,Maltman:2008nf}. For $\Pi_1^{(\prime){\rm OPE}}$, we take into account the terms with dimension $\leq 6$, where the term with dimension=0 ($D$=0) is up to ${\cal O} (\alpha_s^3)$, with $D$=2 (which is proportional to $m_s^2$) up to ${\cal O} (\alpha_s^2)$ and with $D$=4 up to ${\cal O} (\alpha_s^2)$. Note that such radiative corrections for terms can read from \cite{Braaten:1991qm,Chetyrkin:1993hi,Gorishnii:1990vf}. We do not include the radiative correction to the $D$=6 terms since all the uncertainties can be lumped into $a_2$. Note that such radiative corrections for terms with dimensions=0 and 4 are the same as the vector meson case and can read from \cite{Braaten:1991qm,Chetyrkin:1993hi}.

Further applying the Borel (inverse-Laplace) transformation,
 \begin{eqnarray}\label{eq:Borel}
 {\rm\bf B}[f(q^2)]=
\lim_{{\scriptstyle n\to \infty \atop\scriptstyle -q^2\to \infty}
\atop \scriptstyle -q^2/n^2 =M^2 {\rm fixed} }\frac{1}{n!} (-q^2)^{n+1}
\Bigg[{d\over dq^2}\Bigg]^n f(q^2),
 \end{eqnarray}
to both sides of (\ref{eq:dispersion}) to improve the convergence of the OPE series and further suppress
the contributions from higher resonances, the sum rules thus read
\begin{eqnarray}
  f_{f_1}^2 m_{f_1}^2e^{-m_{f_1}^2/M^2}
  &=& \int\limits_0^{s_0^{f_1}}\!\! \frac{s\, ds\,e^{-s/M^2}}{4\pi^2}
  \left[1 + \frac{\alpha_s(\sqrt{s})}{\pi} + F_3 \frac{\alpha_s^2(\sqrt{s})}{\pi^2}
  + (F_4+F_4^\prime \cos^2\theta) \frac{\alpha_s^3 (\sqrt{s})}{\pi^3}
  \right]\nonumber\\
  & &
  -  (\cos\theta -\sqrt{2}\sin\theta)^2
  [\overline{m}_s(\mu_\circ)]^2 \int\limits_0^{s_0^{f_1}}\!\!  ds\,\frac{1}{2\pi^2} e^{-s/M^2}
  \Bigg[1+ \bigg(H_1\ln\frac{s}{\mu_\circ^2}+H_2 \bigg) \frac{\alpha_s(\mu_\circ)}{\pi} \nonumber\\
   & & \ \ \
   + \bigg(H_{3a}\ln^2\frac{s}{\mu_\circ^2}+H_{3b} \ln\frac{s}{\mu_\circ^2} +H_{3c} -\frac{H_{3a}\pi^2}{3}\bigg) \Big(\frac{\alpha_s(\mu_\circ)}{\pi}\Big)^2
  \Bigg]\nonumber\\
  & &
  -\frac{1}{12} \left(1 -\frac{11}{18} \frac{\alpha_s(M)}{\pi}\right)\,
  \langle\frac{\alpha_s}{\pi}G^2\rangle \nonumber\\
  & &
  -\left[\frac{4}{27} \frac{\alpha_s(M)}{\pi}
   +\left( -\frac{257}{486} +\frac{4}{3}\zeta(3) -\frac{2}{27} \beta_1 \gamma_E \right) \frac{\alpha_s^2 (M)}{\pi^2}\right]\,
  \sum_{q_i\equiv u,d,s} \langle \overline{m}_i \bar q_{i} q_{i}\rangle \nonumber\\
  & &
  + \frac{1}{3} (\sqrt{2}\cos\theta +\sin\theta)^2
    \Bigg[
  2 a_1 \overline{m}_{q} \langle \bar qq\rangle
   - \frac{352\pi\alpha_s}{81 M^2} a_2 \langle\bar qq \rangle^2  \Bigg] \nonumber\\
  &  &
  + \frac{1}{3} (\cos\theta -\sqrt{2}\sin\theta)^2
    \Bigg[
  2 a_1 \overline{m}_{s} \langle \bar ss\rangle
   - \frac{352\pi\alpha_s}{81 M^2} a_2 \langle\bar ss \rangle^2  \Bigg]
  \,,\makebox[0.8cm]{}
  \label{eq:SR-f1}
\end{eqnarray}
\begin{eqnarray}
  f_{f_1^\prime}^2 m_{f_1^\prime}^2e^{-m_{f_1^\prime}^2/M^2}
  &=& \int\limits_0^{s_0^{f_1^\prime}}\!\! \frac{s\, ds\,e^{-s/M^2}}{4\pi^2}
  \left[ 1 + \frac{\alpha_s(\sqrt{s})}{\pi} + F_3 \frac{\alpha_s^2(\sqrt{s})}{\pi^2}
  + (F_4+F_4^\prime \sin^2\theta)  \frac{\alpha_s^3(\sqrt{s})}{\pi^3}
  \right]\nonumber\\
  & &
  +  (\sin\theta +\sqrt{2}\cos\theta)^2
  [\overline{m}_s(\mu_\circ)]^2\int\limits_0^{s_0^{f_1^\prime}}\!\!  ds\,\frac{1}{2\pi^2} e^{-s/M^2}
  \Bigg[1+ \bigg(H_1\ln\frac{s}{\mu_\circ^2}+H_2 \bigg) \frac{\alpha_s(\mu_\circ)}{\pi} \nonumber\\
   & & \ \ \
   + \bigg(H_{3a}\ln^2\frac{s}{\mu_\circ^2}+H_{3b} \ln\frac{s}{\mu_\circ^2} +H_{3c} -\frac{H_{3a}\pi^2}{3}\bigg) \Big(\frac{\alpha_s(\mu_\circ)}{\pi}\Big)^2
  \Bigg]\nonumber\\
  & &
  -\frac{1}{12} \left(1 -\frac{11}{18} \frac{\alpha_s(M)}{\pi}\right)\,
  \langle\frac{\alpha_s}{\pi}G^2\rangle \nonumber\\
  & &
  -\left[\frac{4}{27} \frac{\alpha_s (M)}{\pi}
   +\left( -\frac{257}{486} +\frac{4}{3}\zeta(3) -\frac{2}{27} \beta_1 \gamma_E\right) \frac{\alpha_s^2(M) }{\pi^2}\right]\,
  \sum_{q_i\equiv u,d,s} \langle \overline{m}_i \bar q_{i} q_{i}\rangle \nonumber\\
  & &
  + \frac{1}{3} (\sqrt{2}\sin\theta -\cos\theta)^2
    \Bigg[
  2 a_1 \overline{m}_{q} \langle \bar qq\rangle
   - \frac{352\pi\alpha_s}{81 M^2} a_2 \langle\bar qq \rangle^2  \Bigg] \nonumber\\
  &  &
  + \frac{1}{3} (\sin\theta +\sqrt{2}\cos\theta)^2
    \Bigg[
  2 a_1 \overline{m}_{s} \langle \bar ss\rangle
   - \frac{352\pi\alpha_s}{81 M^2} a_2 \langle\bar ss \rangle^2  \Bigg]
  \,,\makebox[0.8cm]{}
  \label{eq:SR-f1p}
\end{eqnarray}
where
\begin{eqnarray}
F_3 &=& 1.9857-0.1153 n_f \simeq 1.6398 \quad {\rm for}\ n_f=3, \nonumber\\
F_4 &=& -6.6368-1.2001n_f-0.0052n_f^2 \simeq -10.2839 \quad {\rm for}\ n_f=3, \nonumber\\
F_4^\prime &=& -1.2395\Delta ,\nonumber\\
H_1 &=& -\frac{8}{81} \beta_1^2 = -2 , \quad
H_2= \frac{2}{9}\beta_2 + 4\beta_2 \Big( \frac{\gamma_1}{\beta_1} - \frac{\gamma_2}{\beta_2} \Big) -\frac{8}{9}\beta_1^2 -4\beta_1\simeq 3.6667 ,\nonumber\\
H_{3a} &=& 4.2499, \quad H_{3b}=-23.1667, \quad H_{3c}= 29.7624, \nonumber\\
 \overline{m}_{q}\langle \bar qq\rangle &\equiv&
 \frac{1}{2}  \left( \overline{m}_{u}\langle \bar uu\rangle + \overline{m}_{d}\langle \bar dd\rangle \right), \quad
 \langle\bar qq \rangle^2 \equiv
 \frac{1}{2}\left( \langle\bar uu \rangle^2 + \langle\bar dd \rangle^2 \right), \nonumber\\
 a_1 &=& 1+ \frac{7}{3}\frac{\alpha_s(M)}{\pi} + \left(\frac{85}{6}-
 \frac{7}{6}\beta_1\gamma_E \right) \frac{\alpha_s^2(M)}{\pi^2},
\end{eqnarray}
with $\beta_1=(2n_f-33)/6$, $\beta_2=(19n_f-153)/12$, $\gamma_1=2, \gamma_2=101/12 -5n_f/18$, and $n_f=3$ being the number of flavors and $\Delta=1$, and $0$ for $f_1$ (singlet) and $f_8$ (octet), respectively \cite{Gorishnii:1990vf}.
In the calculation the coupling constant $\alpha_s(\sqrt{s})$ in Eqs. (\ref{eq:SR-f1}) and (\ref{eq:SR-f1p}) can be expanded in powers of $\alpha_s(M)$:
\begin{eqnarray}
 \frac{\alpha_s(\sqrt{s})}{\pi} &=& \frac{\alpha_s(M)}{\pi}
 +\frac{1}{2} \beta_1 \ln\frac{s}{M^2} \left(\frac{\alpha_s(M)}{\pi} \right)^2
 +\left(\frac{1}{2} \beta_2 \ln\frac{s}{M^2}+ \frac{1}{4} \beta_1^2 \ln^2\frac{s}{M^2}\right) \left(\frac{\alpha_s(M)}{\pi} \right)^3 \nonumber \\
 &+& \left(\frac{\beta_3}{2} \ln\frac{s}{M^2} + \frac{5}{8}\beta_1\beta_2 \ln^2\frac{s}{M^2} + \frac{1}{8} \beta_1^3 \ln^3\frac{s}{M^2} \right) \left(\frac{\alpha_s(M)}{\pi} \right)^4 + \cdots,
\end{eqnarray}
where $\beta_3\simeq -20.1198$. Using the renormalization-group result for the $m_s^2$ term given in \cite{Chetyrkin:1993hi}, we have expanded the contribution to the order ${\cal O}(\alpha_s^2 m_s^2)$ at the subtraction scale $\mu_\circ^2=2$ GeV$^2$ for which the series has better convergence than at the scale 1 GeV$^2$; however, the convergence of the series has no obvious change if using a higher reference scale. As in the case of flavor-breaking $\tau$ decay, the $D=2$ series converges slowly; nevertheless, we have checked that this term, which intends to make the output $m_s$ to be smaller in the fit, is suppressed due to the fact that the mass sum rules for $f_1(1285)$ and $f_1(1420)$ are obtained by applying the differential operator $M^4 \partial\ln /\partial M^2$ to both sides of (\ref{eq:SR-f1}) and (\ref{eq:SR-f1p}), respectively. Nevertheless, the differential operator will instead make the $D$=4 term containing $m_s \langle \bar ss\rangle$ become much more important than the $m_s^2$ term in determining the $f_1(1285)$-$f_1(1420)$ mass difference although the they are the same order in magnitude.

In the numerical analysis, we shall use $\Lambda_{\rm QCD}^{\rm (3)NLO}=0.360$~GeV, corresponding to $\alpha_s(1 {\rm GeV})=0.495$, $\Lambda_{\rm QCD}^{\rm (4)NLO}=0.313$~GeV, and the following values (at the scale $\mu=1$~GeV) \cite{Ioffe:2002be,Narison:2005ny,Ackerstaff:1998yj,Maltman:2008nf}:
\begin{eqnarray}
\begin{array}{l}
  \langle \frac{\alpha_s}{\pi} G_{\mu\nu}^a G^{a\mu\nu} \rangle=(0.009\pm 0.007)~ {\rm GeV}^4 ,\\
  \langle \overline{m}_q\bar qq\rangle= -f_{\pi^+}^2 m_{\pi^+}^2/4\,,\\
  \langle \bar qq \rangle^2  \simeq (-0.247)^6~{\rm GeV}^6 \,,\\
  \langle \bar ss \rangle = (0.30\sim 1.3) \langle \bar qq\rangle \,, \\
    a_2 = -2.9\sim 3.1 \,,
\end{array}\label{eq:parameters}
\end{eqnarray}
where the value of $\langle \bar qq \rangle^2$ corresponds to $(m_u+m_d)(1 {\rm GeV}) \simeq\ 11~ {\rm MeV}$, and  we have cast the uncertainty of $\langle \bar qq \rangle^2$ to $a_2$ in the $D=6$ term.
We do not consider the isospin breaking effect between $\langle \bar uu \rangle$ and $\langle \bar dd \rangle$ since $\langle \bar dd \rangle / \langle \bar uu \rangle -1 \approx -0.007$ \cite{Gasser:1982ap} is negligible in the present analysis. The threshold is allowed by $s_0^{f_1}=2.70\pm0.15$ GeV$^2$ and determined by the maximum stability of the mass sum rule. For an estimate on the threshold difference, we parametrize in the form $(\sqrt{s_0^{f^\prime_1}} -\sqrt{s_0^{f_1}})/\sqrt{s_0^{f_1}} = \delta \times (m_{f_1^\prime}-m_{f_1})/m_{f_1}$, with $\delta =1.0\pm 0.3$. In other words, we assign a 30\% uncertainty to the default value. We search for the allowed solutions for strange quark mass and the singlet-octet mixing angle $\theta$ under the following constraints: (i) Comparing with the observables, the errors for the mass sum rule results of the $f_1(1285)$ and $f_1(1420)$ in the Borel window $0.9$~GeV$^2\leq M^2 \leq 1.3$~GeV$^2$ are constrained to be less than 3\% on average. In this Borel window,
the contribution originating from higher resonances (and the continuum), modeled by
\begin{eqnarray}\label{eq:model-higher-res}
\frac{1}{\pi}\int_{s_0^{f^{(\prime)}}}^\infty ds\, e^{-s/M^2}\, {\rm Im} \Pi_1^{(\prime){\rm
OPE}}(s) \,,
\end{eqnarray}
is about less than 40\% and the highest OPE term (with dimension six) at the quark level is no more than 10\%. (ii) The deviation between the $f_1(1420)-f_1(1285)$ mass difference sum rule result and the central value of the data \cite{PDG} is within $1\sigma$ error: $|(m_{f_1^\prime}-m_{f_1})_{\rm sum \ rule} -144.6~{\rm MeV}|\leq 1.5~ {\rm MeV}$.  The detailed results are shown in Table~1. We also check that if by further enlarging the uncertainties of $s_0^{f_1}$ and $\delta$, $\it e.g.$ 25\%, the changes of results can be negligible. We obtain the strange quark mass with large uncertainty: $m_s (1\, {\rm GeV}) = 106.3 \pm 35.1 ~{\rm MeV}$ (i.e.  $m_s (2\, {\rm GeV}) = 89.5 \pm 29.5 ~{\rm MeV}$) and $\langle \bar{s} s\rangle /\langle \bar{u} u\rangle = 0.56\pm 0.25$ corresponding to $\theta=(19.4^{+4.5}_{-4.6})^\circ$, where the values and $m_s$ and $\langle \bar s s \rangle$ are strongly correlated.

Further accounting for the average of the recent lattice results \cite{McNeile:2010ji,Aoki:2010dy,Durr:2010aw}: $m_s(2~{\rm GeV})=93.6\pm 1.0$~MeV and using the $\theta$ value that we have obtained as the inputs, we get $\langle \bar{s} s \rangle/ \langle \bar{u} u\rangle =0.41 \pm 0.09$ which is less than one and in contrast to the Schwinger-Dyson equation approach in \cite{Williams:2007ef} where the ratio was obtained as $(1.0\pm 0.2)^3$. Our prediction  is consistent with the QCD sum rule result of studying the scalar/pseudoscalar two-point function in \cite{Dominguez:2007hc} where the authors obtained  $\langle \bar{s} s \rangle/ \langle \bar{u} u\rangle = 0.4\sim 0.7$, depending on the value of the strange quark mass.

\begin{table}[t]
\caption{The fitting results in the $f_1(1284)$-$f_1(1420)$ mass difference sum rules. In fit II, we have taken the average of the recent lattice results for $m_s$, which is rescaled to 1 GeV as the input.} \label{tab:result}
\begin{ruledtabular}
\begin{tabular}{ c| c c c c}
~~~~~~~~
    & $m_s (1\, {\rm GeV})$~
    & $\langle \bar{s} s\rangle /\langle \bar{u} u\rangle$
    & $\langle (\alpha_s/\pi) G^2\rangle$
    & $a_2$
 \\
    \hline
Fit I
    & $106.3\pm 35.1$
    & $0.56\pm0.25$
    & $0.0106\pm0.0042$
    & $0.89\pm 0.62$
    \\
    \hline
Fit II
    & [$124.7\pm 1.3]$
    & $0.41\pm0.09$
    & $0.0108\pm0.0037$
    & $0.95\pm 0.45$
\end{tabular}
\end{ruledtabular}
\end{table}
%


\section{Summary}
~~~ We have adopted two different strategies for determining the mixing angle $\theta$: (i) Using the Gell-Mann-Okubo mass formula and the $K_1(1270)$-$K_1(1400)$ mixing angle $\theta_{K_1}=(-34\pm 13)^\circ$ which was extracted from the data for ${\cal B}(B\to K_1(1270) \gamma), {\cal B}(B\to K_1(1400) \gamma), {\cal B}(\tau\to K_1(1270) \nu_\tau)$, and ${\cal B}(\tau\to K_1(1420) \nu_\tau)$, the result is $\theta = (23^{+17}_{-23})^\circ$. (ii) On the other hand, from  the analysis of the ratio of ${\cal B}(f_1(1285) \to \phi\gamma)$ and ${\cal B}(f_1(1285) \to \rho^0\gamma)$, we
have  $\bar\alpha =\theta_i-\theta= \pm(15.8^{+4.5}_{-4.6})^\circ$, i.e., $\theta=(19.4^{+4.5}_{-4.6})^\circ$ or $(51.1^{+4.5}_{-4.6})^\circ$. Combining these two analyses, we deduce the mixing angle $\theta=(19.4^{+4.5}_{-4.6})^\circ$.

We have estimated the strange quark mass and strange quark condensate from the analysis of the $f_1(1420)$-$f_1(1285)$ mass difference QCD sum rule. We have expanded the OPE series up to dimension six, where the term with dimension zero is up to ${\cal O}(\alpha_s^3)$, with dimension=2 up to ${\cal O}(m_s^2 \alpha_s^2)$ and with dimension=4 terms up to ${\cal O}(\alpha_s^2)$. Further using the average of the recent lattice results and the $\theta$ value that we have obtained as the inputs, we get $\langle \bar{s} s \rangle/ \langle \bar{u} u\rangle =0.41 \pm 0.09$.

\section*{Acknowledgments}

This research was supported in part by the National Center for Theoretical Sciences and the
National Science Council of R.O.C. under Grant No. NSC99-2112-M-003-005-MY3.


 \end{document}